\documentclass[showpacs,twocolumn,aps,floatfix]{revtex4}
\usepackage{psfig,epsfig,epsf}
\usepackage{dcolumn}

\newcommand{\beq}{\begin{equation}}  
\newcommand{\eeq}{\end{equation}}  
\newcommand{\bea}{\begin{eqnarray}}  
\newcommand{\eea}{\end{eqnarray}}  

\begin{document}     

\title{Travelling waves in pipe flow}
\author{Holger Faisst}
\email{Holger.Faisst@physik.uni-marburg.de}
\author{Bruno Eckhardt}     
\affiliation{
Fachbereich Physik, Philipps Universit\"at Marburg \\ 
D-35032 Marburg, Germany}    
\date{\today}     
  
\begin{abstract}
A family of three-dimensional travelling waves
for flow through a pipe of circular cross section is identified. 
The travelling waves are dominated by pairs of 
downstream vortices and streaks. 
They originate in saddle-node bifurcations at 
Reynolds numbers as low as $1250$.
All states are immediately unstable. Their dynamical significance 
is that they provide a skeleton for the formation of 
a chaotic saddle that can explain the intermittent transition to 
turbulence and the sensitive dependence on initial
conditions in this shear flow.
\end{abstract}     

\maketitle

Based on decades of studies it is consensus that   
Hagen-Poiseuille flow through a pipe of circular 
cross section belongs to the
class of shear flows that does not become linearly unstable.
Nevertheless, it undergoes an intermittent transition to 
turbulence for sufficiently high Reynolds numbers and
sufficiently large initial perturbations, as first documented
in the classic experiments by Reynolds~\cite{reynolds1883}.
Since then many studies have analyzed the mechanisms of transition
and the properties of the turbulent state~\cite{wygnanski73,eliahou98,ma99,boberg88,brosa99}. 
Particularly relevant to the present analysis are the investigations
by Darbyshire and Mullin~\cite{darbyshire95} which clearly 
show a strong sensitivity to perturbations and a broad intermittent 
range of decaying and turbulent perturbations in an amplitude
vs.~Reynolds number plane. 
Experimental and numerical studies~\cite{eliahou98,schmid01} 
show that the turbulent flow in the 
 transition region 
is 
dominated by downstream vortices
and streaks. 
Various models for their dynamics have been analyzed \cite{trefethen93,holmes96}. 
Taking the full nonlinearity into account Waleffe developed the concept of a nonlinear 
turbulence cycle for the regeneration of vortices and 
streaks~\cite{hamilton95}. 
In addition, stationary solutions and travelling waves have been
found in the full nonlinear equations for 
plane Couette, Taylor-Couette and plane Poiseuille 
flow~\cite{cherhabili96,busse97,waleffe98,FE00,waleffe01}. 
It has been suggested that these structures provide a skeleton for the
transition to turbulence and the observed 
intermittency~\cite{bruno97,bruno02}.
They clearly dominate various observables 
in low Reynolds number turbulent flows~\cite{holmes96}, 
and are also relevant for an understanding of 
the effects of non-Newtonian additives~\cite{waleffe02}.

The existence of exact coherent states in pipe flow 
has been an object of speculation for some time~\cite{boberg88,darbyshire95,brosa99}.
As we will show here Hagen-Poiseuille flow supports
families of travelling waves with structures similar to
those observed in other shear flows as well. This underlines the
significance of vortex-streak interactions also in this system 
and opens alternative routes to modelling and controlling
pipe flow.

The existence of stationary states in plane Couette flow is
connected with an inversion symmetry in the laminar profile.
In the absence of such a symmetry in pipe flow the simplest states we 
can expect are travelling waves (TWs), i.e.~coherent 
structures that move with constant wave speed. The wave speed 
depends on shape and structure and is not known in advance. 
We search for TWs using a Newton-Raphson method applied 
to the full Navier-Stokes equations for an incompressible fluid,
\begin{eqnarray*}
\partial_t \bf u + ({\bf u} \cdot {\bf \nabla}) {\bf u} &=&  
    \nu \Delta {\bf u}  -{ \bf \nabla }p 
\\
\nabla \cdot {\bf u }  &=& 0 \, .
\label{NST}
\end{eqnarray*}
The kinematic viscosity $\nu$ is kept as a free parameter for 
the continuation method and the Reynolds number is determined 
by the mean flow velocity $U$, i.e.{}  
$Re=2RU /\nu$, where $R$ is the pipe radius.
For the numerical representation of the Navier-Stokes equation 
we use a Fourier-Legendre collocation method in cylindrical 
coordinates with Lagrange multipliers to account for  
no-slip boundary conditions at the wall and the constraints
that the flow field is solenoidal, analytical and regular
in the center for $r=0$. The program was verified by reproducing
literature values for the linearized problem~\cite{schmid94},
for the nonlinear dynamics of optimal modes~\cite{zikanov96} 
and for the statistical properties of fully developed turbulent 
flow up to Reynolds numbers of $5000$~\cite{eggels94}.
The Newton-Raphson method was implemented with a spatial 
resolution of $|n/11|+|m/11|<1$, where $n$ and $m$ are the    
azimuthal and downstream wavenumbers, respectively, 
and 56 Legendre polynomials radially. This gives us about $5600$
dynamically active degrees of freedom.

Initial conditions for the Newton method are obtained by embedding
pipe flow into a larger, parameter dependent family of flows.
This has been successful in other cases~\cite{busse97,FE00,waleffe01}. 
For pipe flow Barnes and Kerswell~\cite{barnes00} showed that 
the instabilities that arise from rotation do not extend to the case of 
pure pipe flow.
Guided by the dominance of downstream vortices in the 
stationary and travelling states in planar shear flows we
therefore adopted the following strategy, similar to the one used 
in~\cite{waleffe98}:  
at low Reynolds number a volume force ${\bf F}$ that
generates 
a transverse flow with translation invariant downstream vortices was added. 
While there is no specific physical realization underlying this
volume force, it could be achieved with differential heating and
cooling as predicted by the symmetries of the states, or
by magnetic forcing in an electrically conducting fluid~\cite{brosa91}. 
The amplitude of the driving was then increased beyond the first bifurcation, 
where a stable 3-d TW without translational symmetry appears. 
A path following scheme was then used to track this forced state
through parameter space: the Reynolds number was increased
and the amplitude of the body force decreased. If a 
situation with zero body force could be reached, a 3-d TW for 
the original pipe flow has been found. 

In this way we have identified four different states that 
have a discrete azimuthal
rotation symmetry by construction. An $n$-fold symmetry $C_n$
is defined by the invariance under rotation around the pipe axis 
by an angle $2\pi/n$, i.e.{}  
$$
(u_r,u_\phi,u_z)(r,\phi,z) \rightarrow (u_r, u_\phi, u_z)(r, \phi+2\pi/n, z) \, . 
$$
The resolution in azimuthal direction was adjusted to the
fundamental domains of angle $2\pi/n$ to avoid carrying 
vanishing amplitudes.

\begin{figure}[htb]
\begin{center}
\epsfig{file=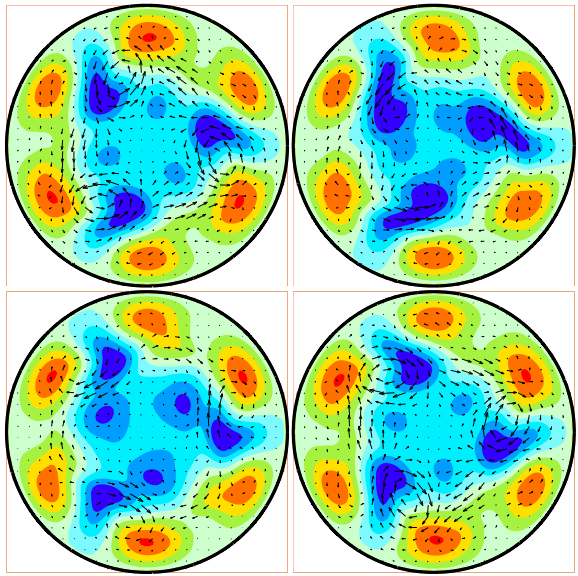,width=.43\textwidth,clip=}
\end{center}
\caption[]{Travelling wave with threefold azimuthal symmetry at the 
bifurcation at $Re=1250$. The frames are cross sections
at different downstream positions separated by $\Delta z=L_z/6$. 
Only half a period is shown: the last frame is the same as the
first one up to a reflection at the horizontal diameter 
($\phi\rightarrow -\phi, u_\phi\rightarrow -u_\phi$).
Note that the six high speed streaks near the wall move much less than
the low speed streaks closer to the center.
Velocity components in the plane 
are indicated by arrows, the downstream component by color coding:
velocities faster than the parabolic profile are shown in green/yellow/red, slower ones
in blue. 
}
\label{TW3}
\end{figure}

The state that bifurcates at the lowest Reynolds number has a three-fold
rotational symmetry in azimuthal direction (Fig.~\ref{TW3}). 
The dynamics of the states is governed by vortices which have a 
predominant downstream orientation 
and are slightly tilted inside the volume. 
They resemble near wall coherent structures in turbulent wall flows
\cite{holmes96}.
They transport slow fluid towards the center 
and fast fluid towards the wall, thus producing high-speed streaks
near the wall and low-speed streaks towards the center (all speeds are
relative to the laminar profile for that Reynolds number).
The steeper gradients near the wall imply higher friction losses
and higher pressure gradients for the same flow speed.
The position of the fast-speed streaks hardly changes in 
downstream direction, whereas the low-speed streaks 
change shape and move in position quite a bit.

In addition to the $C_n$ symmetry externally imposed by the forcing 
the TW's dynamically develop a `shift-and-reflect' symmetry,  
in agreement with the streak instability of other model 
flows~\cite{waleffe95B}: 
if $L_z$ is the wavelength of the state, 
then
\beq
(u_r,u_\phi,u_z)(r,\phi,z) \rightarrow (u_r, -u_\phi, u_z)(r, -\phi, z+L_z/2) \, . 
\label{sr}
\eeq
The same shift-and-reflect symmetry is present in 
3-d coherent states in plane Couette,  
Taylor-Couette and plane Poiseuille flow~\cite{busse97,FE00,waleffe98,waleffe01}.

\begin{figure}[htb]
\begin{center}
\epsfig{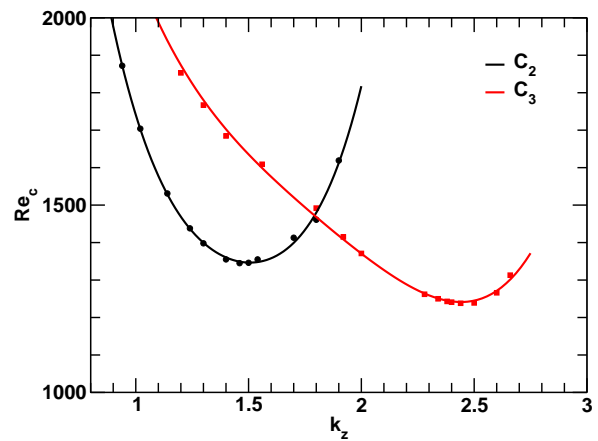}
\end{center}
\caption[]{Dependence of the critical Reynolds number on the 
downstream wavenumber for the two-, three-, and fourfold 
symmetric state. The numerical values (dots) are interpolated
by a 4th-order polynomial fit (continuous lines).
\label{re_L}
}
\end{figure}

Each TW is part of a two-parameter family of states that
exists for a range of Reynolds numbers and a range of downstream
wavelengths (Fig.~\ref{re_L}). 
The lowest critical Reynolds number is obtained for a 
wavelength of about $4.2R$ for the twofold symmetry and about $2.5R$ 
for the three- and fourfold symmetric state. 
This period seems a bit shorter than that
for optimal states in plane Couette flow  where the optimal wavelength
is about $2\pi d$ with $d$ the gap width. However, when the wedge shape
of the boundary is taken into account and the lengths are compared to the
widths, the comparison is more favorable: the ratio widths to 
lengths is about 2$\pi$:4$\pi$ = 1:2 in plane Couette flow 
and about 1:2.5 in pipe flow.

\begin{figure}[htb]
\begin{center}
\epsfig{file=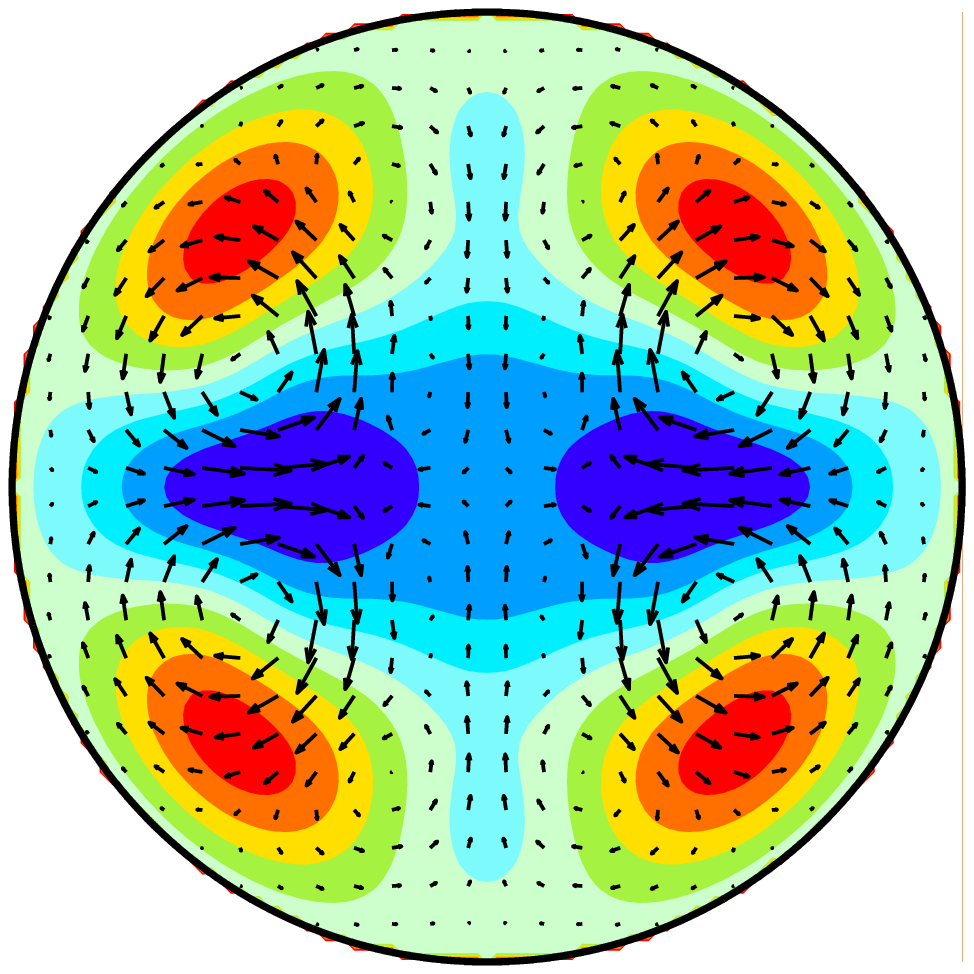,width=.23\textwidth,clip=}
\epsfig{file=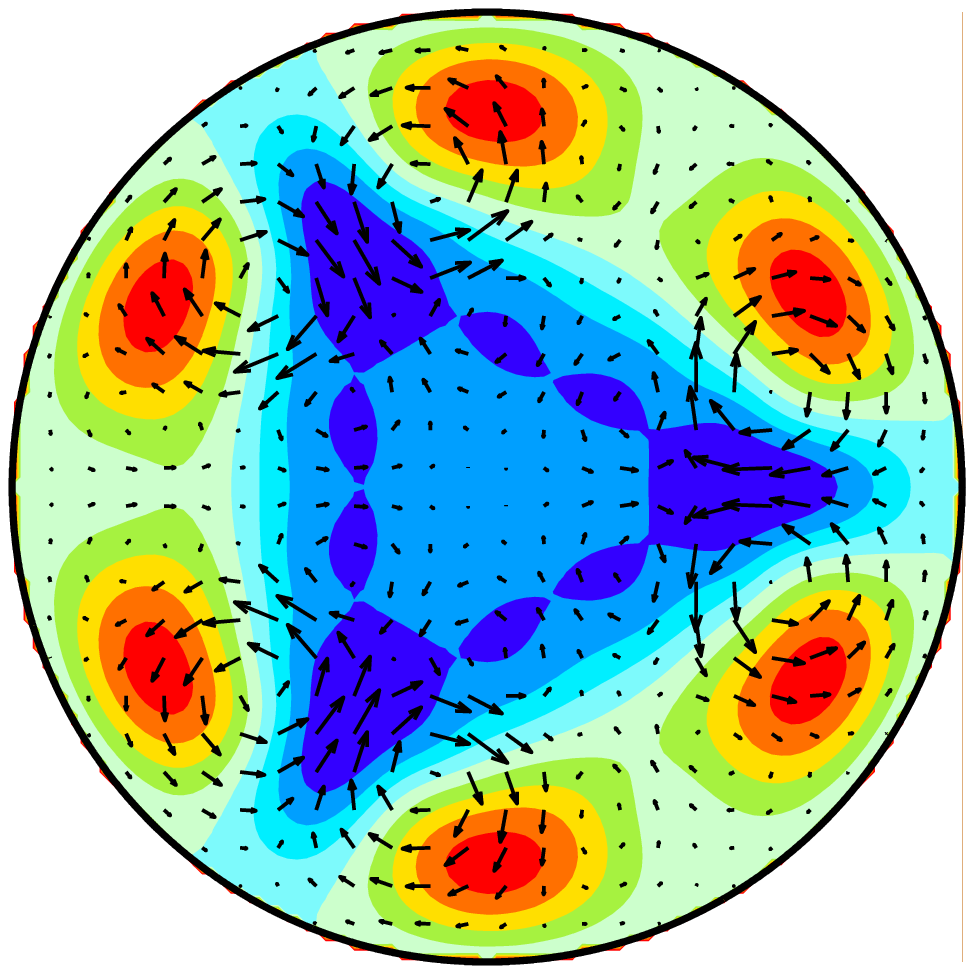,width=.23\textwidth,clip=}
\epsfig{file=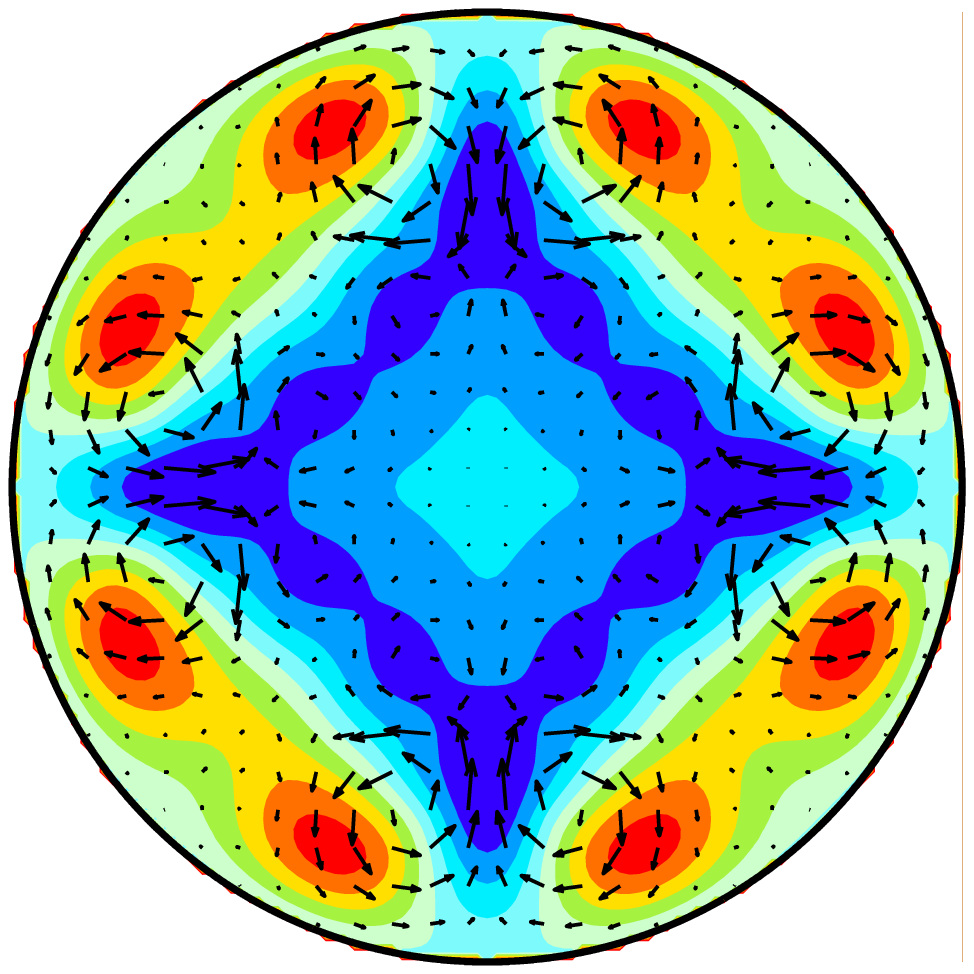,width=.23\textwidth,clip=}
\epsfig{file=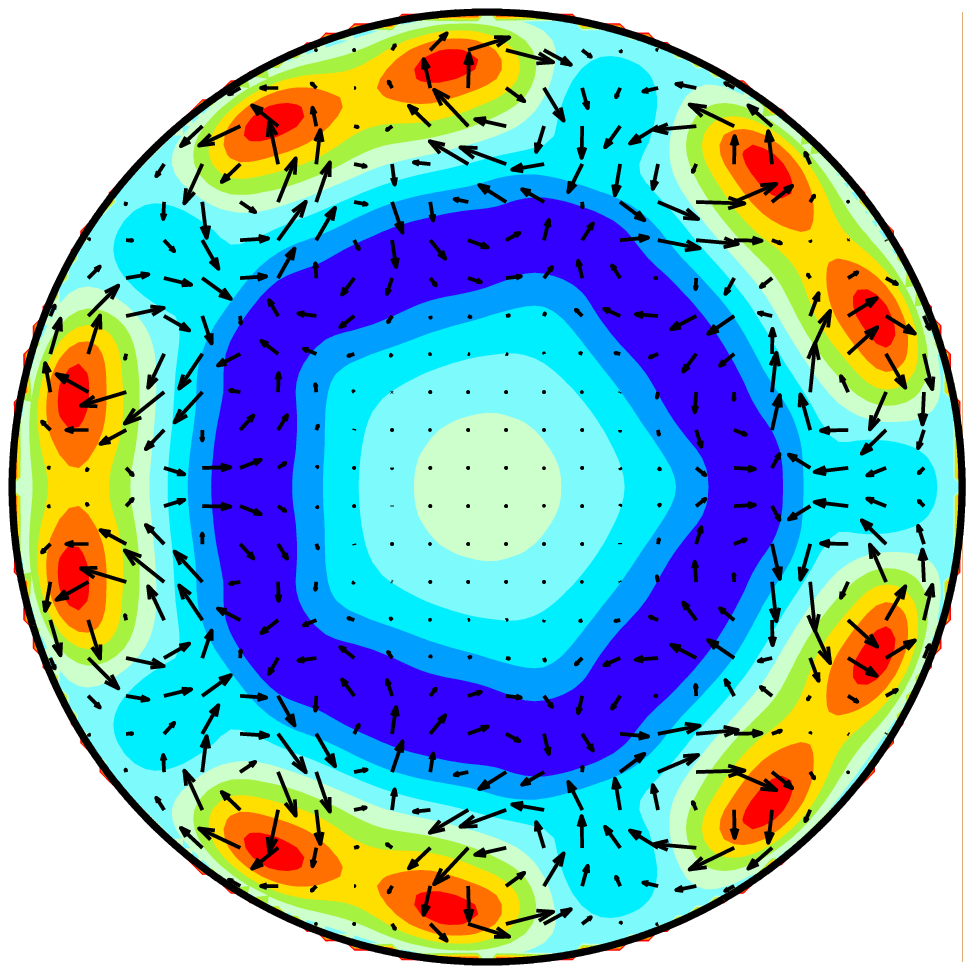,width=.23\textwidth,clip=}
\end{center}
\caption[]{Travelling waves with symmetries $C_n$ with $n=2$, $3$, $4$, and $5$. In order
to highlight the topology of the states all states are averaged in downstream direction. 
The representation of the velocity field by vectors (in-plane motion)
and color (downstream component) is as in Fig.~\ref{TW3}. The absolute scale for the velocity
fields is given in Table~\ref{TW_props}. 
\label{stream_averaged}
}
\end{figure}

Other states that we could identify in pipe flow include 
those with twofold, fourfold and fivefold symmetry (Fig.~\ref{stream_averaged}). 
All these states have a similar topology: they have $2n$ streaks of
fast fluid close to the wall and streaks of slow fluid towards
the center. The high-speed streaks near the wall remain fairly
stationary over one period of the wave, and the low speed streaks
in the center oscillate vigorously.
All states have the shift-and-reflect symmetry (\ref{sr}). 
The wavelength that gives the lowest critical Reynolds number for
these states varies with the 
symmetry (Fig.~\ref{re_L}). 
Several properties of these states are listed in Table~\ref{TW_props}.
The lowest critical Reynolds number is obtained for the TW
with $C_3$ symmetry, for which the arrangement of vortices is
optimal in the sense of being closest to a hexagonal packing,
the preferred pattern in other systems.
The states with $C_4$ and $C_5$ symmetry are not included in the table
as they seem to be the ones which are most 
sensitive to the numerical resolution and not fully converged. 
A preliminary estimate for their critical Reynolds number is about $1590$ for $C_4$ and $2600$ for $C_5$.
\begin{table}[htb]
\begin{tabular}{cddd}
symmetry & \mbox{$C_2$} & \mbox{$C_3$} \\
\hline\hline
$Re_c$  & 1350 & 1250 \\
$L_c/R$ & 4.19 & 2.58 \\
$c/
{
U
}$  & 1.43 & 1.29 \\
$n_u$ & 2 & 1 \\
$u_s/
{
U
}$ & 0.38  & 0.35   \\
$u_t/
{
U
}$ & 0.035 & 0.046    
\end{tabular}
\caption[]{Selected properties of travelling waves at the saddle-node bifurcation.
Given is the critical Reynolds number $Re_c$ at the optimal wavelength $L_c$, 
the phase velocity $c$ and the number $n_u$ of unstable dimensions.
$u_s$ is the maximum deviation of the streamwise velocity from the laminar flow, 
$u_t$ is the maximum transverse velocity component.
}
\label{TW_props}
\end{table}

Interestingly, the list of TWs does not include a state with the $C_1$-
symmetry of the mode that shows the strongest linear transient 
growth~\cite{bergstroem93,schmid94}. 
A state with $C_1$ symmetry could be found for 
non-vanishing volume force, but up to Reynolds numbers of $4000$ it could
not be continued to pure pipe flow without reconnecting to the 
laminar profile. This should have implications for
low dimensional model building~\cite{bergstroem99,brosa99} 
where linear arguments have led to an emphasis of the $C_1$ mode.

\begin{figure}[htb]
\begin{center}
\epsfig{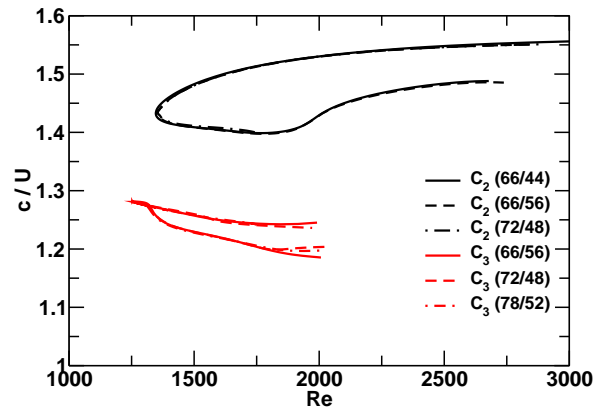}
\end{center}
\caption[]{Phase velocities for the travelling states at optimal wavelength
as a function of Reynolds number. The velocities are normalized by the mean
downstream velocity for the states. 
Three different resolutions are plotted, 
indicated by ($J/K$) where  $J$ is the 
number of independent Fourier modes, $K$ is the number of Legendre polynomials.  
\label{phaser}
}
\end{figure}

The phase speed $c$ of the TWs is a function of both
parameters, length and Reynolds number. 
In all cases the phase speed is larger than the mean speed but slower
than the maximal speed possible with a laminar fluid (Fig.~\ref{phaser}).
That is to say, the wave still propagates downstream when viewed from a frame of 
reference moving with the mean flow velocity.

By advecting the unperturbed profile small transverse
components can produce strong downstream streaks. 
Table~\ref{TW_props} shows that
for the TWs the transverse
components are about an order of magnitude smaller than the 
differences between the downstream components of the laminar profile 
and the TW.

The TWs appear as finite amplitude saddle-node bifurcations.
Away from the bifurcations there is an upper (nodal)
and a lower (saddle) branch. The saddle has one more positive
eigenvalue than the node, but in all cases we found that even
at the point of bifurcation at least one eigenvalue is positive:
all states are thus unstable right from the point of creation.
This is in contrast to plane Couette flow, where at least the lowest
state has an interval of stability, although a tiny one
\cite{busse97}.

\begin{figure}[htb]
\begin{center}
\epsfig{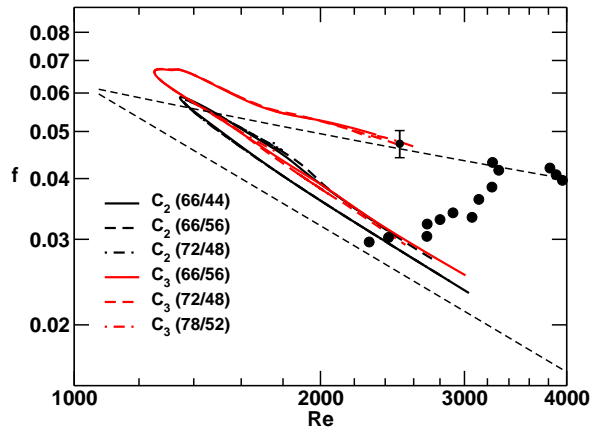}
\end{center}
\caption[]{Travelling waves and turbulent states in pipe flow.
The friction factor as defined in Eq.~(\ref{ff}) is shown 
vs.~Reynolds number. The TWs
are labeled by their azimuthal symmetries.
Three different resolutions are plotted as explained in Fig.~\ref{phaser}. 
The lower dashed line indicates the strict lower bound from 
the laminar profile (where $f= 64/Re$), 
the upper dashed line gives the mean friction factor 
from a logarithmic profile.
The full circles are experimental data 
from~\cite{schlichting79} or from our numerical simulations
(the value at $Re=2500$ with error bars).
\label{friction}
}
\end{figure}
The friction properties of all states 
together with experimental and numerical results for 
turbulent flows are shown in Fig.~\ref{friction}.
The friction factor $f$ is defined as 
\beq
f = {4 R\Delta p}/{\rho U^2 L_z} \, ,
\label{ff}
\eeq
with $\Delta p$ the pressure drop over a distance $L_z$ 
along the pipe and $\rho$ the density
of the fluid.
The saddle-node bifurcations take place at Reynolds numbers significantly below
the values where typical perturbations induce turbulent dynamics. 
At the bifurcation the friction factor is higher than the value extrapolated
from turbulent states at higher $Re$.
The upper branches in Fig.~\ref{friction} correspond to the 
lower branches in Fig.~\ref{phaser}: states with 
higher friction have lower phase velocity and vice versa.

Besides the TWs presented here we can expect many other TWs.
On the assumption that narrower vortices will require stronger
driving we anticipate that they will have critical
Reynolds number higher than the $1250$ found for the $C_3$ state. 
In addition, there are all the states that bifurcate from the TWs, 
as well as states that bifurcate in secondary bifurcations. 
Furthermore, there will be homoclinic and heteroclinic connections 
between these TWs as well as other periodic and aperiodic 
states. All these states together then form a chaotic saddle 
in phase space that supports the turbulent dynamics~\cite{tel}.
One characteristic of this chaotic saddle is an exponential distribution
of life times and this is indeed found in numerical 
simulations~\cite{pipe_trans}. It will be interesting to see 
to which extent these states can be used
to approximate properties of the turbulent flow. 

Support by the Deutsche Forschungsgemeinschaft is greatfully acknowledged.


\bibliographystyle{plain}

\begin{thebibliography}{99}  

\bibitem{reynolds1883}
O.~Reynolds.
\newblock { Phil. Trans. R. Soc.} {\bf 174}, 935 (1883).

\bibitem{wygnanski73}
I.J.~Wygnanski{,} F.H.~Champagne.
\newblock { J. Fluid Mech.} {\bf  59}, 281 (1973).
I.J.~Wygnanski{,} M.~Sokolov{,} D.~Friedman.
\newblock { J. Fluid Mech.} {\bf 69}, 283 (1975).

\bibitem{eliahou98}
S.~Eliahou{,} A.~Tumin{,} I.~Wygnanski.
\newblock { J. Fluid Mech.} {\bf 361}, 333 (1998).
G.~Han{,} A.~Tumin{,} I.~Wygnanski.
\newblock { J. Fluid Mech.} {\bf 419}, 1 (2000).
H.~Shan{,} B.~Ma{,} Z.~Zhang{,} F.T.M.~Nieuwstadt.
\newblock { J. Fluid Mech.} {\bf 387}, 39 (1999).

\bibitem{ma99}
B.~Ma{,} C.W.~van~Doorne{,} Z.~Zhang{,} F.T.M.~Nieuwstadt.
\newblock { J. Fluid Mech.} {\bf 398}, 181 (1999).

\bibitem{boberg88}
L.~Boberg{,} U.~Brosa.
\newblock { Z. Naturforsch.} {\bf 43a}, 697 (1988).

\bibitem{brosa99}
U.~Brosa{,} S.~Grossmann.
\newblock { Eur. Phys. J. B} {\bf  9}, 343 (1999).

\bibitem{darbyshire95}
A.G.~Darbyshire{,} T.~Mullin.
\newblock { J. Fluid Mech.} {\bf 289}, 83 (1995).

\bibitem{schmid01}
P.J.~Schmid{,} D.S.~Henningson.
\newblock{Stability and {T}ransition in {S}hear {F}lows},
\newblock{Springer, (2001)}

\bibitem{trefethen93}
L.N.~Trefethen{,} A.E.~Trefethen{,} S.C.~Reddy{,} T.A.~Driscol.
\newblock { Science} {\bf 261}, 578 (1993).
S.~Grossmann.
\newblock { Rev. Mod. Phys.} {\bf 72}, 603 (2000).

\bibitem{holmes96}
P.~Holmes{,} J.L.~Lumley{,} G.~Berkooz.
\newblock {Turbulence, {C}oherent {S}tructures, {D}ynamical {S}ystems and {S}ymmetry}.
\newblock Cambridge {U}niversity {P}ress, (1996).

\bibitem{hamilton95}
J.M.~Hamilton{,} J.~Kim{,} F.~Waleffe.
\newblock { J. Fluid Mech.} {\bf 287}, 317 (1995).
F.~Waleffe.
\newblock { Phys. Fluids} {\bf 7}, 3060 (1995).
F.~Waleffe.
\newblock { Phys. Fluids} {\bf 9}, 883 (1997).

\bibitem{cherhabili96}
A.~Cherhabili{,} U.~Ehrenstein.
\newblock { J. Fluid Mech.} {\bf 342}, 159 (1997).

\bibitem{busse97}
M.~Nagata;
\newblock { J. Fluid Mech.} {\bf 217}, 519 (1990).
R.M.~Clever{,} F.H.~Busse.
\newblock { J. Fluid Mech.} {\bf 344}, 137 (1997).
M.~Nagata.
\newblock { Phys. Rev. E} {\bf 55}(2), 2023 (1997).

\bibitem{waleffe98}
F.~Waleffe.
\newblock { Phys. Rev. Lett.} {\bf 81},  4140 (1998).

\bibitem{FE00} 
H.~Faisst{,} B.~Eckhardt.
\newblock { Phys. Rev. E} {\bf 61}, 7227 (2000).

\bibitem{waleffe01}
F.~Waleffe.
\newblock { J. Fluid Mech.} {\bf 435}, 93 (2001).

\bibitem{bruno97}
A.~Schmiegel{,}~B. Eckhardt.
\newblock { Phys. Rev. Lett.} {\bf 79}, 5250 (1997).

\bibitem{bruno02}
B.~Eckhardt{,} H.~Faisst{,} J. Schumacher and A.~Schmiegel.
\newblock { Advances in turbulence {IX}}, 
\newblock I.P. Castro, P.E. Hanock and T.G. Thomas (eds), Barcelona, 701 (2002)

\bibitem{waleffe02}
P.A.~Stone{,} F.~Waleffe{,} M.D.~Graham.
\newblock { Phys. Rev. Lett.} {\bf 89}, 208301 (2002).

\bibitem{schmid94}
P.J. Schmid{,} D.S.~Henningson.
\newblock { J. Fluid Mech.} {\bf  277}, 197 (1994).

\bibitem{zikanov96}
O.Y. Zikanov.
\newblock { Phys. Fluids}  {\bf 8}, 2923 (1996).

\bibitem{eggels94}
J.G.M.~Eggels{,} F.~Unger{,} M.H.~Weiss{,} J.~Westerweel{,} R.J.~Adrian{,} R.~Friedrich{,} F.T.M.~Nieuwstadt.
\newblock { J. Fluid Mech.} {\bf  268}, 175 (1994).
M.~Quadrio{,} S.~Sibilla.
\newblock { J. Fluid Mech.} {\bf 424}, 217 (2000).


\bibitem{barnes00}
D.R. Barnes{,}~R.R. Kerswell.
\newblock { J. Fluid Mech.} {\bf 417}, 103 (2000).

\bibitem{brosa91}
U.~Brosa.
\newblock { Z. Naturforsch. \bf 46a}, 473 (1991).

\bibitem{waleffe95B}
F.~Waleffe.
\newblock { Stud. Appl. Math.} {\bf 95}, 319 (1995).

\bibitem{bergstroem93}
L.~Bergstr\"om.
\newblock { Phys. Fluids A} {\bf 5}, 2710 (1993).

\bibitem{bergstroem99}
L.~Bergstr\"om.
\newblock { Eur. J. Mech. B} {\bf 18}, 635 (1999).

\bibitem{schlichting79}
H.~Schlichting.
\newblock {Boundary-{L}ayer {T}heory}.
\newblock New {Y}ork: Mc{G}raw-{H}ill, (1979).

\bibitem{tel}
T.~T\'el.
\newblock {\em Directions in Chaos}, 
Vol. 3, ed. Hao, B.-L. (World Scientific, Singapore) (1990), pp. 149-221.

\bibitem{pipe_trans}
H.~Faisst{,} B.~Eckhardt.
\newblock{ Sensitive dependence on initial conditions 
in transition to turbulence in pipe flow},
\newblock{ to be published}


\end{thebibliography}


\end{document}